\documentclass[letterpaper,11pt]{article}
\usepackage{jheppub_mod}
\usepackage{graphicx, amsmath, amssymb, amstext}
\usepackage{relsize, float, multirow}
\usepackage{array, epstopdf, hyperref, }
\usepackage{caption, subcaption, mathtools}
\usepackage{newtxmath}
\usepackage[normalem]{ulem}

\title{Constraints on $N_{\rm{eff}}$ of high energy non-thermal neutrino injections upto $z\sim 10^8$ from CMB spectral distortions and abundance of light elements  }
\author[a]{Sandeep Kumar Acharya,} 
\author[a]{Rishi Khatri}
\affiliation[a]{Department of Theoretical Physics, Tata Institute of 
Fundamental Research, Mumbai 400005, India}
\emailAdd{sandeepkumar@theory.tifr.res.in, khatri@theory.tifr.res.in}
\date{\today}
\abstract{ 
High energy neutrinos and anti-neutrinos ($\gtrsim$ 100 GeV) can inject energetic electromagnetic particles  into the baryon-photon plasma in the high redshift universe through electroweak showers from electroweak bremsstrahlung, inelastic scattering with the background electrons and nucleons, and by pair-production of standard model particles on background neutrinos and anti-neutrinos. In this paper, we evolve the particle cascades of high energy non-thermal neutrinos injections, using dark matter decay as a specific example, including relevant collision processes of these neutrinos with the background particles and taking into account the  expansion of the universe. We study the effect of these non-thermal neutrino injections on the CMB spectral shape and abundance of light elements produced in the big bang nucleosynthesis. We show that CMB spectral distortions and abundance of light elements can constrain neutrino energy density at the recombination, parameterized as contribution to $N_{\rm{eff}}$, from high energy neutrino injection. These constraints are stronger by several orders of magnitudes compared to the CMB anisotropy constraints. We also show that CMB spectral distortions can probe neutrino injections to significantly higher redshifts ($z>2\times 10^6$) as compared to pure electromagnetic energy injection.   
}
\notoc
\begin{document}
\maketitle
\newpage
\section{\label{sec:intro}Introduction}
Precise measurement of the cosmic microwave background anisotropy (CMB) \cite{Pl2018} has not only established the standard cosmological model with 6 parameters but it has also  allowed us to study extensions of the standard model with extra parameters. One such extension is the energy density in free streaming relativistic particles usually parameterized by effective number of neutrino species ($N_{\rm{eff}}$) defined using the CMB energy density as reference. The current 2-$\sigma$ constraint on $N_{\rm{eff}}$ from CMB anisotropy is 2.92$\pm$0.36 \cite{Pl2018}. The standard model prediction for $N_{\rm{eff}}$ is 3.046 \cite{DT1992,HM1995,DHS1997,GG1998,DHS1999,D2002,MMPP2002,MMPPPS2005,BYR2015,DS2016,
GDP2019}, for a more recent calculation see \cite{AY2020}. The extra relativistic energy density affects the expansion rate of the Universe. This makes $N_{\rm{eff}}$ somewhat degenerate with the Hubble parameter. By requiring the angular measure of the acoustic scale at the recombination epoch to be fixed (which is precisely measured by \emph{Planck} \cite{Pl2018}), constraints on the Hubble parameter lead to constraints on the total radiation energy density of the universe at the recombination epoch i.e. on the $N_{\rm{eff}}$ parameter. The extra neutrino species which are thermalized, in addition to the standard model neutrinos, or other new relativistic particles such as dark radiation can make $N_{\rm{eff}}>3.046$ while addition of extra energy to photons after neutrino decoupling can decrease the relative energy density in neutrinos with respect to the CMB and hence make $N_{\rm{eff}}<3.046$. However, we need not restrict ourselves to such scenarios involving new particles. We will consider the case where the extra $N_{\rm{eff}}$ is contributed by the energetic standard model neutrinos. One example is high energy non-thermal neutrino injections from dark matter decay in the pre-recombination era.  \par
\hspace{1cm}
 In almost all previous studies, it is assumed that the standard model interactions of neutrinos with the  background particles are too weak to be important for any cosmological signature with the exception of their effect on BBN \cite{D2002}. While this may be true of low energy neutrinos, energetic neutrinos above the W, Z boson mass scale can emit electromagnetic particles through electroweak showers from electroweak bremsstrahlung \cite{BK02002,KS2007,CCRSSU2011}, inelastic scattering of injected neutrinos with the background nucleons (proton and helium nuclei) and electrons \cite{GQHS1996,CMS2011,FZ2012}, and pair production of standard model particles by annihilation of injected neutrinos/anti-neutrinos with background anti-neutrinos and neutrinos \cite{GGS1993,R1993}. If the injected neutrino's (anti-neutrino's) energy is sufficiently high then it can pair produce electron-positron pairs and quark pairs on background anti-neutrinos (neutrinos). The quarks after hadronization and pion decay produce secondary electromagnetic particles and neutrinos. These electromagnetic particles, in the pre-recombination era, can produce CMB spectral distortion or modify the abundance of primordial elements. \par
 \hspace{1cm}
 Electromagnetic energy injection into the background baryon-photon plasma modifies the Planckian CMB spectrum, creating a distortion in the spectral shape. Assuming this distortion to be created by interaction of non-relativistic thermal electrons, heated by energy injection to the baryon-photon fluid, energy injection at $z\lesssim 2\times 10^6$ leads to $y,i,$ and $\mu$-type ($yim$ collectively) spectral distortions \cite{Sz1969,Sz19701,Is19752,Bdd1991,Chluba:2011hw,Ks2012b,Chluba:2013vsa}. Recently, it was shown that at $z\lesssim 2\times 10^5$, spectral distortion shapes (non-thermal relativistic or $ntr$-type) from high energy photon/electron injection can be substantially different from $yim$-type distortions \cite{AK2018,AK20191}. High energy photon/electron injection gives rise to a particle cascade. Particles in the time-evolving cascade can have relativistic energy which give rise to richer spectral distortion shapes as compared to $yim$-distortions. For energy injection at $z\gtrsim 2\times  10^5$, $ntr$-spectral distortions also thermalize to $\mu$-distortion \cite{AK20191}. While photon production processes such as bremsstrahlung and double Compton scattering are inefficient at  $z\lesssim 2\times 10^6$, their increasing efficiency above this redshift, in the thermalization epoch or blackbody photosphere,   wash out any spectral distortion from electromagnetic energy injection and create a Planck spectrum with modified temperature. The transition from $\mu$ distortion to a temperature shift is captured by the $\mu$-visibility function \cite{Sz19701,Bdd1991,ks2012,C2014}. The weakness of neutrino interactions implies that they deposit their energy gradually. In particular,  neutrinos injected at $z\gtrsim 2\times 10^6$ will deposit some of their energy in the photon-baryon plasma at $z\lesssim 2\times 10^6$ and leave imprints in CMB spectral distortions. We thus have a new window into the thermalization epoch where any signature of direct electromagnetic energy injection is wiped out from the CMB. In general, for any energy injection process at $z\gtrsim 2\times 10^6$, the electromagnetic part of the injected energy will be thermalized and will be invisible, except as contribution to the CMB energy density or a change in $N_{\rm{eff}}$,  while a fraction of energy injected in neutrinos will be deposited later and will be visible as CMB spectral distortions.  \par
 \hspace{1cm}    
 Injection of energetic photons above the photo-dissociation threshold of deuterium and helium can change the abundance of primordial elements. Given the precise measurements of abundance of elements, subject to astrophysical uncertainties, we can constrain electromagnetic energy injection upto redshifts $z\sim 3\times 10^6$ \cite{ENS1985,EGLNS1992,KM1995,PS2015,KKMT2018,HSW2018,FMW2019,AK20192}. The energetic photons can dissociate nuclei or can scatter with background electrons through Compton scattering, scatter with the CMB photons elastically, or pair produce electrons and positrons. Pair-production on the CMB photons is the dominant process when it is kinematically allowed. Since the energy of CMB photons increases with redshift ($\propto (1+z)$), the threshold energy of injected photons to pair-produce $e^-e^+$ decreases with redshift. When the pair-production threshold on the CMB photons is below helium photo-dissociation threshold of 19.81 MeV, most of the injected photons immediately pair-produce $e^-e^+$ and the pair-produced particles are not energetic enough to dissociate helium nuclei. This happens at $z\sim 3\times 10^6$ and the constraints from photo-dissociation of BBN elements become weaker at $z\gtrsim 3\times 10^6$. We can probe higher redshifts with deuterium destruction (photo-dissociation threshold=2.2 MeV), but constraints from deuterium destruction are much weaker due to its low abundance \cite{KM1995}. \par
 \hspace{1cm} 
    A sufficiently energetic neutrino above the quark pair production threshold releases $\approx$ 50 percent of its energy in electromagnetic particles \cite{CCHHKPRSS2011} and the rest as secondary neutrinos. Assuming these secondary neutrinos do not pair produce, they will lose some of their energy to the baryon-photon plasma through various scattering processes and their remaining energy can be parameterized by $N_{\rm{eff}}$. The electromagnetic energy deposition at $2\times 10^5 \lesssim z\lesssim 2\times 10^6$ gives rise to $\mu$-distortions and the energy in $\mu$ distortion is constrained to be smaller than $\approx 10^{-4}$ of the CMB energy density ($\mu\lesssim 9\times 10^5$) \cite{F1994,Cobe1994,F1996}. This means that $N_{\rm{eff}}$ from high energy neutrino injections is also constrained to be of the order of $10^{-4}$ of the CMB energy density, which is many orders magnitude stronger than the current CMB anisotropy constraints. The CMB anisotropy constraints are of course more general while the CMB spectral distortion constraints are very model dependent and apply only to neutrinos with energy $\gtrsim$ 100 GeV. Similar analysis can also be done for modification to the abundance of primordial elements.
\par
\hspace{1cm}    
     For neutrinos with energy below the pair-production threshold of quarks with background anti-neutrinos but above the W, Z boson mass also, CMB spectral distortions and abundance of light elements can give reasonably stronger constraints compared to \emph{Planck}, as we will show. For CMB spectral distortion constraints many orders of magnitude improvements is possible with future experiments such as Primordial Inflation Explorer (PIXIE) \cite{FM2002,Pixie2011}. \par
    \hspace{1cm}
    Modification to the abundance of primordial elements from high energy neutrino injection was previously considered in \cite{GSS1991,DS1993} (see also \cite{NS2014}). However, their analysis was limited to neutrino energy of 1 GeV- $10^3$ GeV. Also, electroweak shower from neutrinos due to electroweak bremsstrahlung and inelastic scattering with background nucleons and electrons were not considered. In this paper, we consider neutrino injection from 10 GeV to $10^{12}$ GeV from dark matter decay. Detection of PeV neutrinos at \emph{IceCube} \cite{Icecube2013} has inspired many particle physics models \cite{CGIT2010,FKMY2013,ABGHMDW2015,HKKM2018} in which dark matter can decay to high energy neutrinos. We consider monochromatic neutrino injection and evolve the neutrino spectrum from the redshift of energy injection upto the recombination epoch, taking into account the aforementioned scattering processes in the expanding universe. Since we are considering small effects on CMB and abundance of elements, results for any general neutrino spectrum can be obtained by linear superposition of the monochromatic results. We will calculate the constraints on the  fraction of decaying dark matter and the resulting $N_{\rm{eff}}$, obtained from CMB spectral distortions and abundance of BBN elements, as a function of dark matter decay lifetime or the redshift of energy injection.  
    \section{Evolution of neutrino cascade in the expanding universe}
    \begin{figure}[!tbp]
  \begin{subfigure}[b]{0.4\textwidth}
    \includegraphics[scale=0.8]{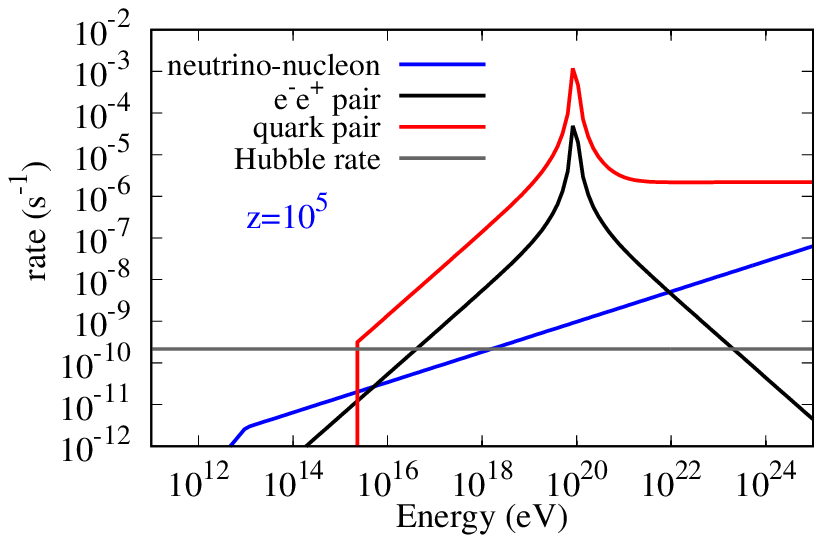}
    \caption{}
    \label{fig:ratez10^5}
  \end{subfigure}
  \hfill
  \begin{subfigure}[b]{0.4\textwidth}
    \includegraphics[scale=0.8]{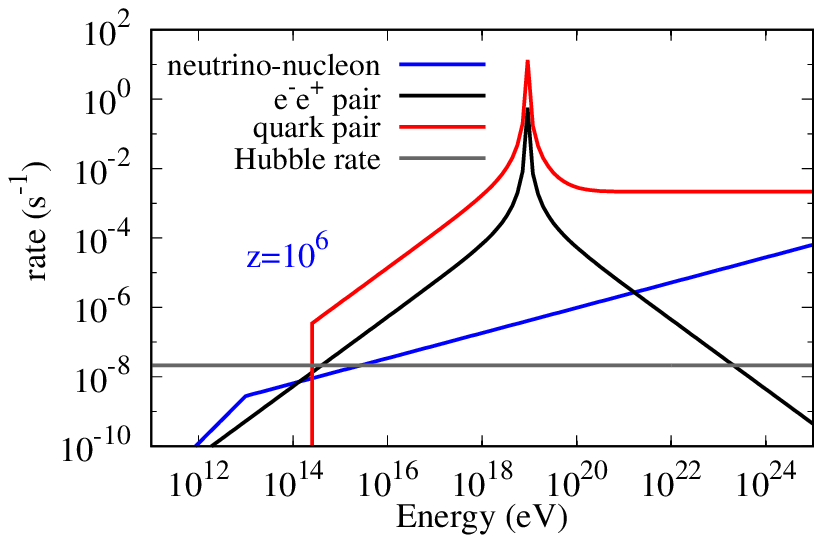}
    \caption{}
    \label{fig:ratez10^6}
  \end{subfigure}
  \caption{Collision rates of neutrinos or antineutrinos as a function of incident neutrino or antineutrino energy at (a)$z=10^5$ (b)$z=10^6$. The peak corresponds to the energy of the incident neutrino such that the center of mass energy for annihilation with background neutrino or anti-neutrino is equal to the Z-boson mass.}
  \label{fig:rate}
\end{figure}
\begin{figure}
\begin{center}
    \includegraphics[scale=1.0]{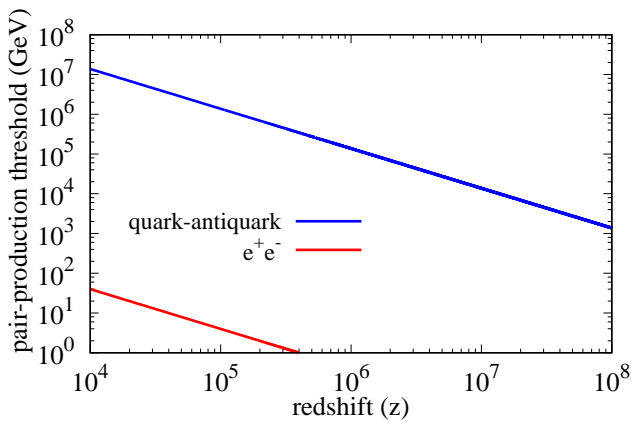}
    \caption{Threshold energy of injected neutrinos (anti-neutrinos) to pair-produce quark-antiquark and electron-positron on annihilation with the background anti-neutrinos (neutrinos). }
    \label{fig:pairp}
    \end{center}
  \end{figure}
We evolve the injected neutrino distribution in the expanding universe as they interact with the background particles and follow the subsequent particle cascade. We consider neutrino-nucleon scattering and pair production on background neutrino (antineutrinos) to produce electron-positron pairs and quark-gluon pairs. The comparison of the collision rates of these scattering processes with the Hubble rate is shown in Fig. \ref{fig:rate} at two redshifts. The cross-section for neutrino-nucleon or neutrino-electron scattering is proportional to the square of center of mass energy which is proportional to the mass of the target particle in the energy range of interest. Therefore, scattering on electrons is negligible compared to nucleons \cite{FZ2012}. The difference between the cross-sections for inelastic scattering of high energy neutrinos and anti-neutrinos with nucleons is negligible (See Eq. 14 of \cite{GQHS1996}). There is no difference between pair-production of standard model particles from high energy neutrinos or anti-neutrinos by annihilation with the background anti-neutrinos or  neutrinos respectively since in the center of mass frame they are indistinguishable. The cross-section of neutrinos with the nucleons can be parameterized as a broken power law \cite{GQHS1996}. The break in the power law is due to the increasing importance of scattering with the sea quarks as compared to the valence quarks inside the nucleons. The fraction of neutrino's initial energy retained by the neutrino is parameterized as the inelasticity parameter (see Table 1 and 2 of \cite{GQHS1996}). This can vary from 50 percent for 10 GeV neutrinos to 20 percent for $10^{12}$ GeV neutrinos. If the energy of the incident neutrinos is high enough, they can pair produce standard model particles on annihilation with the background anti-neutrinos. The interaction can have both neutral and charged current contribution \cite{GGS1993,R1993}. The peak in Fig. \ref{fig:rate} is the energy of the incident neutrino such that the center of mass energy for neutrino-antineutrino collision is equal to the Z-boson mass. Beyond this energy, the contribution of neutral current to total cross-section dies away while the contribution of the charged current process dominates the total cross-section \cite{R1993}. We consider all quarks and gluons to have a mass threshold of 300 MeV, which is the kinematical mass of up and down quarks \cite{PYTHIA2006,PYTHIA2015}, above which the quarks and gluons are freely emitted.  The cross-section for quark pair production is more than an order of magnitude higher compared to $e^-e^+$ pair production due to more degrees of freedom of quarks and gluons \cite{GGS1993}. Therefore, beyond the incident neutrino energy at which quark pair production starts (center of mass energy $\approx$ 300 MeV), it is the most dominant process. The charged current contribution, important only at high energies, to $e^-e^+$ pair production channel, is therefore not important and we neglect it. In Fig. \ref{fig:pairp}, we plot the threshold energy of injected neutrinos/anti-neutrinos for pair-production of quark-antiquark and electron-positron pairs by annihilation with background antineutrinos/neutrinos. \par
\hspace{1cm}
The quarks after hadronization produce pions which decay to secondary electrons, positrons, photons, neutrinos, and stable hadrons. The authors in \cite{CCHHKPRSS2011} have provided spectra of these decay products from injection of any standard model particle with center of mass energy in the range 5 GeV-$10^6$ GeV. This center of mass energy can be translated to lab energy of incident neutrino as $E_{\nu}=\frac{1}{4}s_{\rm{CM}}/E_{\bar{\nu}}$, where $s_{\rm{CM}}$ is the square of center of mass energy and $E_{\bar{\nu}}$ is the energy of background anti-neutrinos. The energy of background neutrinos and anti-neutrinos at $z\sim 10^8$ is 20 keV.  Therefore, the result of \cite{CCHHKPRSS2011} can be used for very high energy incident neutrinos upto energy of $E_{\nu}=10^{17}$ GeV in the cosmological setup. While the spectrum provided in \cite{CCHHKPRSS2011} is in the center of mass (CM) frame, we have to boost it to the CMB frame. For this, we assume the distribution of secondary electromagnetic particles and neutrinos in the center of mass frame to be spherically symmetric. The square of center of mass energy for collisions of injected high energy neutrino with energy $E_{\nu}$ with  background anti-neutrino is given by,
\begin{equation}
s_{\rm{CM}}=4.0 \times E_{\nu}\times (\frac{4}{11})^{1/3} k_{\rm{B}}T_{\rm{CMB}},
\end{equation}
where $k_{\rm{B}}$ is the Boltzmann constant and $T_{\rm{CMB}}=2.725(1+z)$K is the CMB temperature. We assume that the background neutrinos and anti-neutrinos to have energy$ \approx k_{\rm{B}}T_{\nu}\approx$ $(\frac{4}{11})^{1/3} k_{\rm{B}}T_{\rm{CMB}}$, where $T_{\nu}$ is the temperature of neutrinos in standard $\Lambda$CDM cosmology. If the neutrinos are their own antiparticles (Majorana) the number density of each flavor $i$ of neutrino at redshift $z$ is given by, 
\begin{equation}
n_{\nu}^i(z)\equiv n_{\bar{\nu}}^i(z)=(\frac{3}{4})\times (\frac{4}{11}) n_{\gamma}(z),
\label{neunumdens}
\end{equation}
where $n_{\gamma}$ is the total number density of CMB photons at redshift $z$. The background number density is important only for the annihilation of injected neutrinos/antineutrinos with the cosmic  neutrino background (CNB) particles. For Dirac particles, the number density of the target background neutrinos and anti-neutrinos  gets halved. However, whenever the $\nu\bar{\nu}$ annihilation is important compared to the other collision processes, it is also much faster compared to the Hubble rate (Fig. \ref{fig:rate}). This means that almost all injected neutrinos would annihilate in this regime and a factor of 2 change in the number density of the annihilation targets due to neutrinos being Dirac or Majorana fermions has a negligible effect on the results. Our results are therefore valid irrespective of whether neutrinos are Dirac or Majorana fermions.  
\par
\hspace{1cm}
The Lorentz factor of center of mass frame is given by, $\gamma_{\rm{CM}}=\frac{E_{\nu}}{\sqrt{s_{\rm{CM}}}}$. A monochromatic energy source with spherically symmetric distribution in the CM frame is boosted to a box distribution in the lab frame or Friedmann metric/background frame i.e.,
\begin{equation}
\frac{dN}{dE}\propto \Theta(E-E_-)\Theta(E_+-E),
\end{equation}
where $N$ is the comoving number density of particles, $\Theta(x)$ is the Heaviside step function such that $\Theta(x)=1$ for $x>0$ and $\Theta(x)=0$ for $x<0$, $E_{\pm}=\frac{\gamma_{\rm{CM}}}{2}E(1 \pm \beta_{\rm{CM}})$, $\beta_{\rm{CM}}$ is the boost factor, and $E$ is the energy of secondary neutrinos in the center of mass frame.
We superpose the lab spectrum for all secondary neutrinos produced in quark hadronization to get the full spectrum. Although the secondary neutrinos will have lower energies compared to the original injected neutrinos, some secondary neutrinos in the high energy tail of the spectrum can still pair-produce. Low energy neutrinos mostly redshift while depositing a small fraction of their energy due to inelastic scattering with nucleons. The electromagnetic energy deposited at $z\gtrsim 2\times 10^5$ thermalizes to a Bose-Einstein spectrum and $\mu$-type distortion is a good approximation \cite{AK2018}. In particular, in this regime of thermal spectral distortions, only total injected electromagnetic energy is required and the actual spectrum of injected electromagnetic  particles is not needed. However, we need the spectrum of electromagnetic particles in order to calculate the photo-dissociation of light elements produced in the BBN since there is a threshold energy below which electromagnetic particles are unable to photo-dissociate the elements. Therefore, we calculate the spectra of electromagnetic particles as well. These secondary electromagnetic particles can lead to electromagnetic cascades which themselves have to be evolved in the expanding universe \cite{AK2018}, but in this work we ignore this. Instead, we assume that any electromagnetic energy injected at a particular timestep is deposited at that redshift as heat to the background electrons giving rise to CMB spectral distortion or goes into photo-dissociation of nuclei. We,  therefore, calculate the photo-dissociation of elements only for the initial injected photons in the on-the-spot approximation, which is a good approximation at $z\gtrsim 10^5$. We compare the on-the-spot approximation with the full calculation in Appendix \ref{app:bbn}.\par
\hspace{1cm}
A fraction of neutrinos, with energy greater than the mass of W, Z bosons, can decay to these particles which after hadronic and leptonic cascade produce stable standard model particles on timescales much shorter than the Hubble time \cite{CCRSSU2011}. The particle spectrum with electroweak shower  can be computed in \textbf{PYTHIA} \cite{PYTHIA2015} and has been included in the data provided by \cite{CCHHKPRSS2011}. Thus, neutrinos (anti-neutrinos) with energy $\sim$ 100 GeV can still give rise to reasonable spectral distortion and BBN signature, even though they can not pair produce quarks on the background anti-neutrinos (neutrinos). \par
\hspace{1cm}
To evolve the particle spectrum, we divide the energy $\sim$ eV- $10^{21}$ eV in 300 log spaced dimensionless energy bins in dimensionless energy variable  $x=\frac{E}{\rm{k_B T_{\rm{CMB}}}}$, where $\rm{k_B}$ is the Boltzmann constant and $T_{\rm{CMB}}$ is the CMB temperature. Using dimensionless  energy variable $x$ takes care of redshifting of energy due to expansion of the universe  implicitly. A particle only redshifting away its energy will stay in the same $x$ bin. The details of our calculation of particle cascade are described in Appendix \ref{app:algo}. 
\section{Dark matter decaying into neutrinos and the resulting $N_{\rm{eff}}$}
 We consider high energy monochromatic neutrino injection from decay of long-lived unstable particles in the dark sector. The energy injection from dark matter decay can be parameterized by,
\begin{equation}
\frac{d\rho}{dt}=\frac{f_X}{\tau_X}\rho_c c^2 (1+z)^3 e^{-(t/\tau_X)},
\end{equation}
where $\rho$ is the injected energy density, $f_X$ is the fraction of decaying dark matter compared to total dark matter, $\rho_c$ is the energy density of total dark matter, $\rm{c}$ is the speed of light, $\tau_X$ is the dark matter lifetime with corresponding redshift denoted as $z_X$. The relative change in the CMB energy density due to electromagnetic energy injection in a redshift bin $|\Delta z|$ at $z$ is given by,
\begin{equation}
\Delta \epsilon(z)=\frac{\Delta \rho_{\rm{CMB}}(z)}{\rho_{\rm{CMB}}(z)},
\end{equation}
 where $\Delta \rho_{\rm{CMB}}$ is the injected electromagnetic energy density and $\rho_{\rm{CMB}}$ is the standard model CMB energy density given by $\rho_{\rm{CMB}}=0.26(1+z)^4$ eV/cm$^3$. The total fractional change of CMB energy density at a particular redshift $z$ is given by integration of fractional change of CMB energy density from starting redshift of energy injection ($z_{st}$) upto that redshift i.e.,
\begin{equation}
\epsilon(z)=\int_{z_{st}}^z \frac{d[\Delta \rho_{\rm{CMB}}(z_{inj})/\rho_{\rm{CMB}}(z_{inj})]}{dz_{inj}} dz_{inj}.
\label{epsilon}
\end{equation}
The fractional change of CMB energy density which shows up as spectral distortion in the CMB is given by,
\begin{equation}
\epsilon_{SD}(z)=\int_{z_{st}}^z \frac{d[\Delta \rho_{\rm{CMB}}(z_{inj})/\rho_{\rm{CMB}}(z_{inj})]}{dz_{inj}}\rm{e} ^{-\tau(z_{inj})}dz_{inj},
\label{epsilonSD}
\end{equation}
where $\tau(z_{inj})$ is the $\mu$-visibility function, which is given by $\tau(z_{inj})\approx \rm{e}^{(-(z_{inj}/z_\mu)^{2.5})}$ with $z_\mu=2\times 10^6$ \cite{Sz19701,ks2012,C2014}.
  The amplitude of $\mu$-type distortion from energy injection can be written as,
\begin{equation}
\mu=1.4\int_{z_{st}}^{z_t}\frac{d[\Delta \rho_{\rm{CMB}}(z_{inj})/\rho_{\rm{CMB}}(z_{inj})]}{dz_{inj}}\rm{e} ^{-\tau(z_{inj})}dz_{inj},
\label{totalmu}
\end{equation} 
where $z_t\approx 2\times 10^5$. At $z\lesssim z_t$, $\mu$-type distortion is no longer a good approximation and we should take into account the actual spectrum of spectral distortions \cite{AK20191} when calculating constraints.
 The energy density of relativistic freestreaming particles is parameterized by $N_{\rm{eff}}$ such that their energy density well after electron-positron annihilation is given by
\begin{equation}
\rho_{\nu}=N_{\rm{eff}}\left(\frac{7}{8}\right)\left(\frac{4}{11}\right)^{4/3}\rho_{\rm{CMB}},
\end{equation}
   where $\rho_{\nu}$ is the energy density of neutrinos in $\Lambda CDM$ cosmology and $N_{\rm{eff}}=3.046$. In case of electromagnetic energy injection, $\rho_{\rm{CMB}}$ will increase compared to $\rho_{\nu}$ and $N_{\rm{eff}}$ will be smaller compared to the $\Lambda CDM$ model. On the other hand, if we inject neutrinos, the $N_{\rm{eff}}$ will be expected to increase if the energy of neutrinos $\lesssim$ GeV. For the case of ultra-high energy neutrinos that we are interested in, with energy $\gtrsim 100$ GeV, a fraction of initial energy injected in neutrinos will be lost to electromagnetic particles and end up in increasing the energy density of CMB. Depending upon what fraction of initial neutrino energy is dissipated electromagnetically and what fraction remains with secondary neutrinos, $N_{\rm{eff}}$ can either increase or decrease. \footnote{A related discussion can be found in Sec. 6.1 of \cite{CRA2020} and \cite{SS2008}.} With energy injection, the expression for modified $N_{\rm{eff}}$ is given by,
\begin{equation}
N_{\rm{eff}}^*=\frac{\rho_{\nu}^*}{\left(\frac{7}{8}\right)\left(\frac{4}{11}\right)^{4/3}\rho_{\rm{CMB}}^*},
\end{equation}
where $\rho_{\nu}^*=\rho_{\nu}+\Delta \rho_{\nu}$, $\rho_{\rm{CMB}}^*=\rho_{\rm{CMB}}+\Delta \rho_{\rm{CMB}}$, where $\Delta \rho_{\nu}$ and $\Delta \rho_{\rm{CMB}}$ are the change in neutrino and CMB energy density due to energy injection respectively.   
For small energy injections ($\frac{\Delta \rho_{\rm{CMB}}}{\rho_{\rm{CMB}}}$,$\frac{\Delta \rho_{\nu}}{\rho_{\nu}}<<1$), the expression for $N_{\rm{eff}}^*$ can be written as,
\begin{equation}
N_{\rm{eff}}^*=N_{\rm{eff}}+N_{\rm{eff}}\left(\frac{\Delta \rho_{\nu}}{\rho_{\nu}}-\frac{\Delta \rho_{\rm{CMB}}}{\rho_{\rm{CMB}}}\right),
\end{equation}
and the deviation in free streaming relativistic degrees of freedom $\Delta N_{\rm{eff}}=N_{\rm{eff}}^*-N_{\rm{eff}}$ is given by,
\begin{equation}
\Delta N_{\rm{eff}}=N_{\rm{eff}}\left(\frac{\Delta \rho_{\nu}}{\rho_{\nu}}-\frac{\Delta \rho_{\rm{CMB}}}{\rho_{\rm{CMB}}}\right).
\label{delNeff}
\end{equation}
Total $\Delta N_{\rm{eff}}(z)$ at $z$ from energy injection at all higher redshifts can be calculated by summing up contribution from all previous redshift bins as,
\begin{equation}
\Delta N_{\rm{eff}}(z)=\int_{z_{st}}^z \frac{ d(\Delta N_{\rm{eff}})}{dz_{inj}} d z_{inj}.
\label{toalNeff}
\end{equation}  
The electromagnetic energy from high energy neutrinos contribute to $\Delta \rho_{\rm{CMB}}$ while the surviving primary as well as secondary neutrinos contribute to $\Delta \rho_{\nu}$. The electromagnetic energy released before $z=2\times 10^6$ thermalizes creating a temperature shift of the CMB while energy deposited at later redshifts create a $ntr$ or $\mu$ distortion. We note that even when electromagnetic energy injection does not create spectral distortions at $z\gtrsim 2\times 10^6$, it still changes $N_{\rm{eff}}$ and contributes to $\Delta N_{\rm{eff}}$ i.e. both temperature shift and distortion in the CMB contribute to $\Delta \rho_{\rm{CMB}}$. \par
\hspace{1cm}
\begin{figure}[!tbp]
  \begin{subfigure}[b]{0.4\textwidth}
    \includegraphics[scale=0.8]{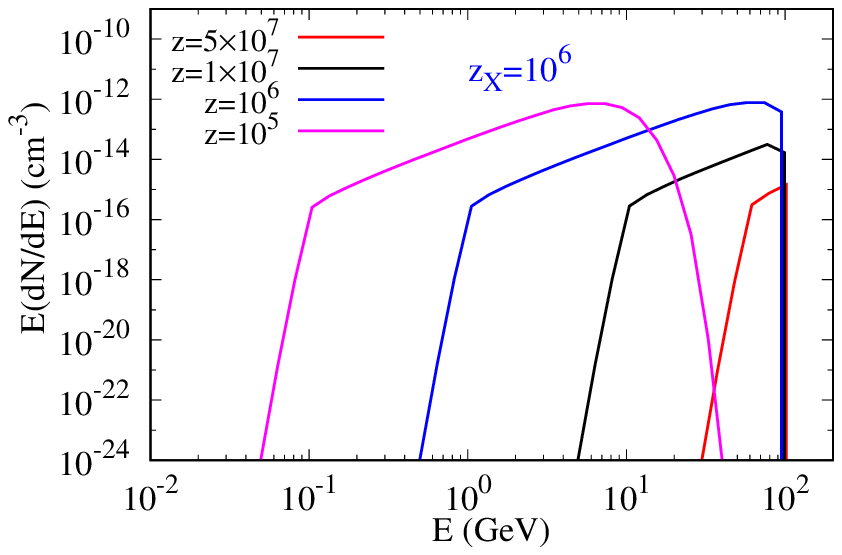}
    \caption{neutrino}
    \label{fig:nspectrum10^2gevzx10^6}
  \end{subfigure}
  \hfill
  \begin{subfigure}[b]{0.4\textwidth}
    \includegraphics[scale=0.8]{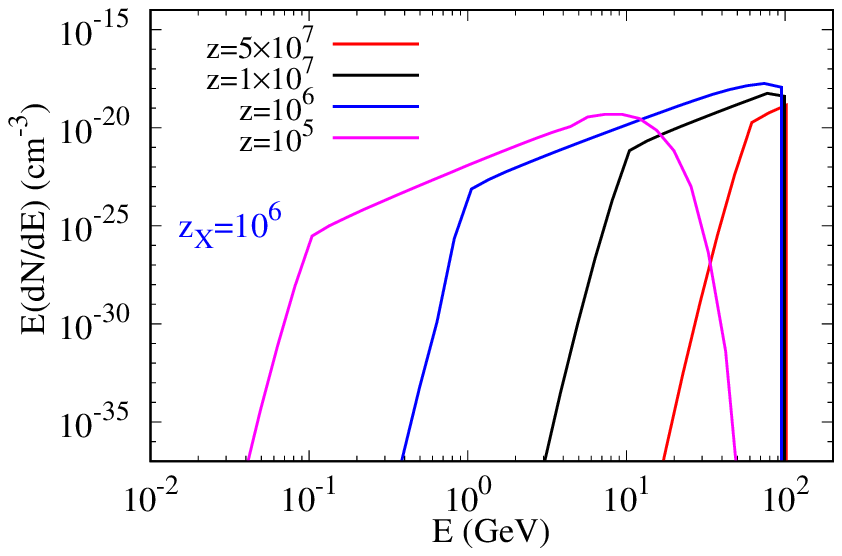}
    \caption{Photon/electron}
    \label{fig:EMspectrum10^2gevzx10^6}
  \end{subfigure}
  \begin{subfigure}[b]{0.4\textwidth}
    \includegraphics[scale=0.8]{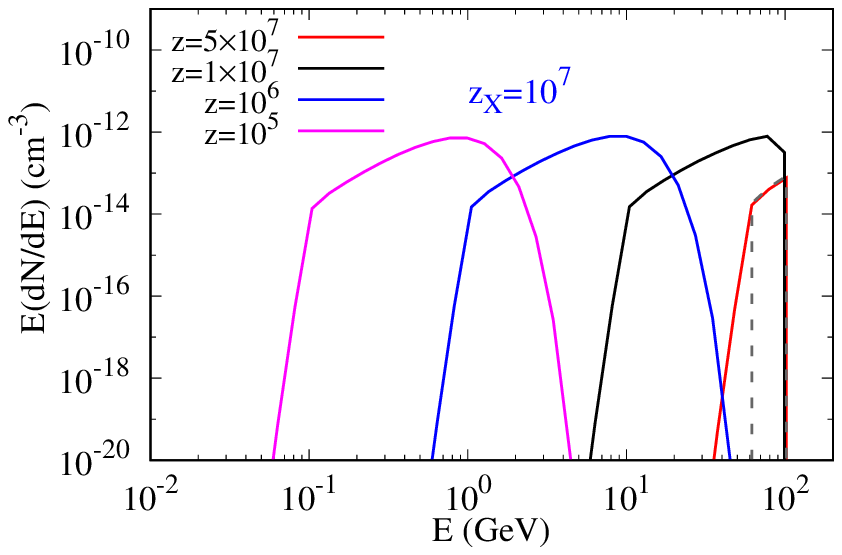}
    \caption{neutrino}
    \label{fig:nspectrum10^2gevzx10^7}
  \end{subfigure}
  \hfill
  \begin{subfigure}[b]{0.4\textwidth}
    \includegraphics[scale=0.8]{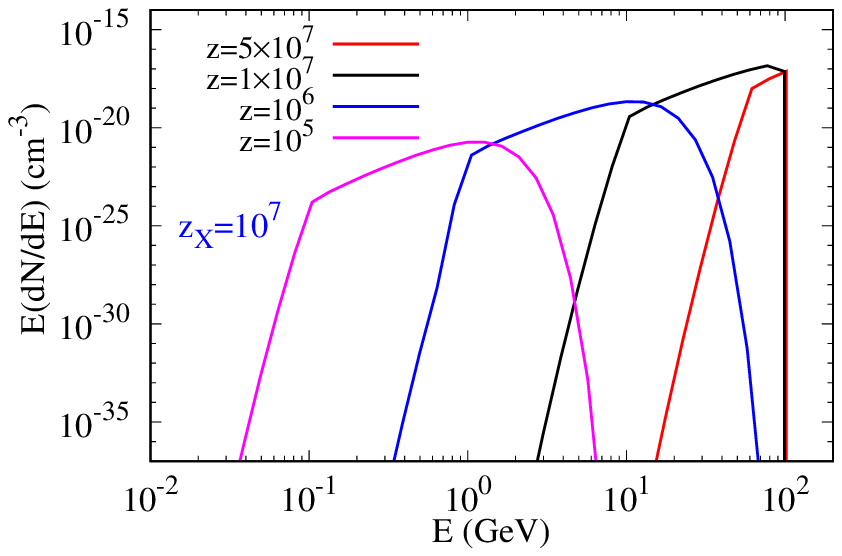}
    \caption{Photon/electron}
    \label{fig:EMspectrum10^2gevzx10^7}
  \end{subfigure}
    \caption{Instantaneous spectrum of neutrinos and electromagnetic particles $\frac{dN}{d\rm{ln}E}$ (comoving energy density per unit log energy) at different redshifts for 100 GeV initial neutrino injection from dark matter decay with lifetime as denoted in the figure. The electromagnetic spectrum is not evolved in redshift but obtained from instantaneous neutrino spectrum at that redshift. The dotted line in \ref{fig:nspectrum10^2gevzx10^7} is obtained by neglecting neutrino-nucleon scattering. We have assumed $f_X=0.1$ for this figure.  }
    \label{fig:spectrum10^2GeV}
  \end{figure}
  \begin{figure}[!tbp]
  \begin{subfigure}[b]{0.4\textwidth}
    \includegraphics[scale=0.8]{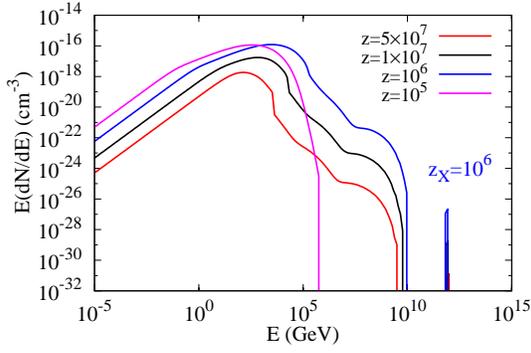}
    \caption{neutrino}
    \label{fig:nspectrum10^12gevzx10^6}
  \end{subfigure}
  \hfill
  \begin{subfigure}[b]{0.4\textwidth}
    \includegraphics[scale=0.8]{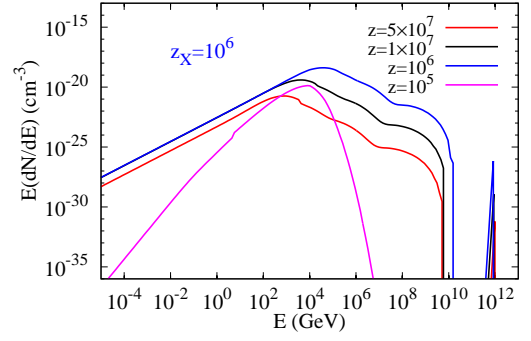}
    \caption{Photon/electron}
    \label{fig:EMspectrum10^12gevzx10^6}
  \end{subfigure}
  \begin{subfigure}[b]{0.4\textwidth}
    \includegraphics[scale=0.8]{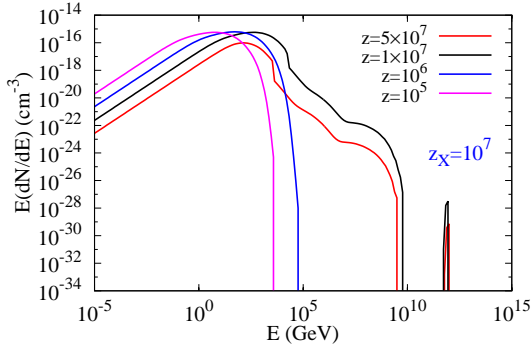}
    \caption{neutrino}
    \label{fig:nspectrum10^12gevzx10^7}
  \end{subfigure}
  \hfill
  \begin{subfigure}[b]{0.4\textwidth}
    \includegraphics[scale=0.8]{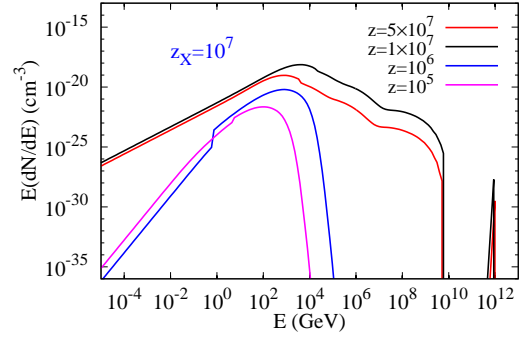}
    \caption{Photon/electron}
    \label{fig:EMspectrum10^12gevzx10^7}
  \end{subfigure}
    \caption{Instantaneous spectrum of injected neutrinos and electromagnetic particles $\frac{dN}{d\rm{ln}E}$ (comoving number density per unit log energy) at different redshifts for $10^{12}$ GeV initial neutrino injection from dark matter decay with lifetime as denoted in the figure. We have assumed $f_X=0.1$ for this figure.  }
    \label{fig:spectrum10^12GeV}
  \end{figure}
\begin{figure}[!tbp]
  \begin{subfigure}[b]{0.4\textwidth}
    \includegraphics[scale=0.8]{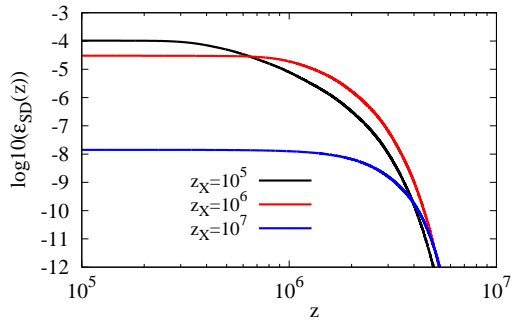}
    \label{fig:yT}
  \end{subfigure}
  \hfill
  \begin{subfigure}[b]{0.4\textwidth}
    \includegraphics[scale=0.8]{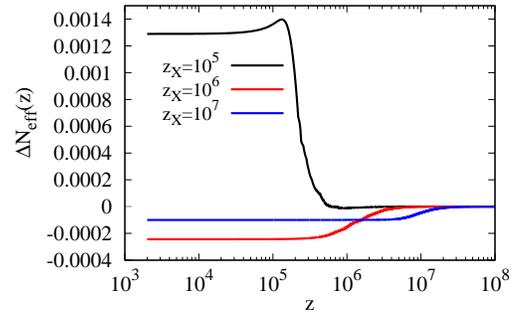}
    \label{fig:NeffT}
  \end{subfigure}
  \caption{ (a) $\epsilon_{SD}(z)$ and (b) $\Delta N_{\rm{eff}}$ from injection of $10^6$ GeV neutrinos as a function of redshift with $f_X=10^{-4}$ for different $z_X$.}
\label{fig:yNeffT}
  \end{figure}
  In Fig. \ref{fig:spectrum10^2GeV} and \ref{fig:spectrum10^12GeV}, we plot the instantaneous neutrino and electromagnetic spectrum of injected primary as well as secondary particles for dark matter decay with different $z_X$ for initial injection of 100 GeV and $10^{12}$ GeV neutrinos. For $10^{12}$ GeV neutrino injection, there are surviving neutrinos at $10^{12}$ GeV as well as secondary neutrinos after quark-antiquark pair production which have significantly lower energy. For 100 GeV neutrinos, there is no quark-antiquark pair-production and we have a smooth and continuous spectrum from surviving neutrinos. The instantaneous electromagnetic spectrum is obtained from energy release of instantaneous neutrino spectrum after collision with background particles. The electromagnetic particles as well as neutrinos are produced after hadronization and electromagnetic energy is released from neutrino-nucleon scattering which is a function of neutrino spectrum. Therefore, the spectrum of electromagnetic particles is qualitatively similar to neutrinos. In Fig. \ref{fig:nspectrum10^2gevzx10^7}, the dotted curve is calculated ignoring neutrino-nucleon scattering. With neutrino-nucleon scattering taken into account, the spectrum spreads out a little as neutrinos lose energy.   
  \par
  \hspace{1cm}
In Fig. \ref{fig:yNeffT}, we show the evolution of $\epsilon_{SD}(z)$ and $\Delta N_{\rm{eff}}(z)$ by energy injection from dark matter decay (with $f_X=10^{-1}$) to $10^6$ GeV neutrinos for a few values of $z_X$. The starting redshift for calculation ($z_{st}$) is chosen to be $10^8$. At higher redshifts, the probability for neutrinos to release their energy as electromagnetic energy is higher because the background neutrinos and anti-neutrinos are at higher temperature, and hence, there is a higher probability of electron-positron and quark-anti-quark pair production. There can be multiple cycles of pair productions as high energy tails of secondary neutrinos can again pair-produce on background anti-neutrinos and vice versa. Also, there is a higher probability of neutrino-nucleons inelastic scattering as the density of nucleons (or baryons) is higher at higher redshifts. Therefore, contribution to $\Delta N_{\rm{eff}}(z)$ from fractional change in CMB energy density dominates and $\Delta N_{\rm{eff}}(z)$ is negative. At lower redshifts, extra neutrino energy density contribution to $\Delta N_{\rm{eff}}(z)$ starts to dominate as there is an increase in the probability for neutrinos to survive. Both $\epsilon_{SD}(z)$ and magnitude of $\Delta N_{\rm{eff}}(z)$ decrease with increasing $z_X$ as energy release from dark matter decay is proportional to $(1+z)^3$ while the background neutrino and CMB energy density $\propto$ $(1+z)^{4}$. Once the dark matter has completely decayed, there is no more energy injection and $\epsilon_{SD}(z)$ and $\Delta N_{\rm{eff}}(z)$ become constant as interaction of surviving low energy neutrinos with background particles becomes negligible. The decrease in magnitude of $\epsilon_{SD}(z)$ is more drastic compared to $\Delta N_{\rm{eff}}(z)$ with increasing redshift due to the $\mu$-visibility function. We have assumed that the amount of energy injection with respect to the CMB is small. This assumption is not valid at high redshifts since at $z>2\times 10^6$ energy injection of order $\rho_{\rm{CMB}}$ is also allowed by the COBE data because of the exponential suppression of spectral distortions. A recent study \cite{CRA2020} has relaxed this assumption at least for one-time energy injection. For continuous energy injection, we must calculate on case-by-case basis. We will, however, continue to use the simple analytic expression for the $\mu$-visibility function since we are mostly interested in  small spectral distortions created by surviving neutrinos at $z\lesssim 2\times 10^6$. The increase in the efficiency of photon production processes such as bremsstrahlung and double Compton scattering results in any electromagnetic energy injection to thermalize, creating a Planck spectrum with higher temperature and exponential decrease in the magnitude of the CMB spectral distortion at $z\gtrsim 2\times 10^6$. There is no such restriction for neutrinos since they interact weakly, deposit their energy slowly, and can survive for a long time. In particular, neutrinos injected at $z\gtrsim 2\times 10^6$ can survive until $z\lesssim 2\times 10^6$ and deposit some of their energy in electromagnetic particles at a time when the CMB spectral distortions can survive. \par
\hspace{1cm}
\begin{figure}[!tbp]
  \begin{subfigure}[b]{0.4\textwidth}
    \includegraphics[scale=0.8]{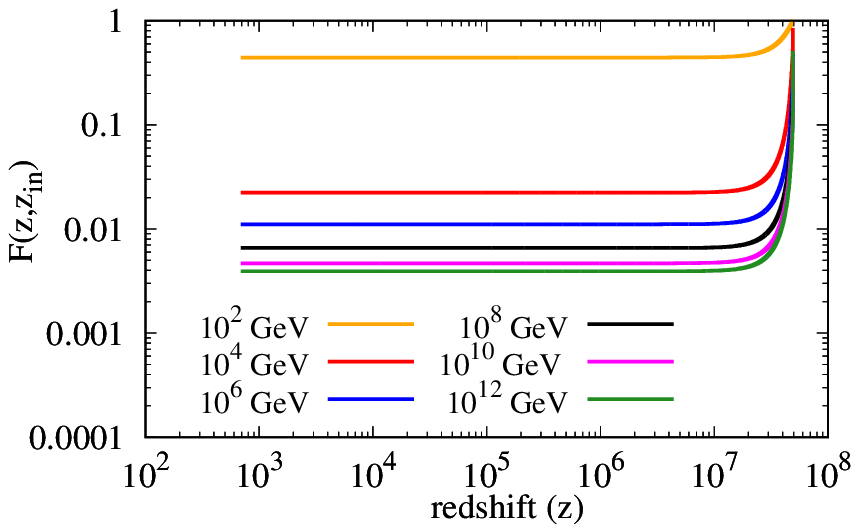}
    \caption{Fraction of surviving neutrino energy $F(z,z_{inj})$) as a function of redshift for one-time energy injection at $z_{inj}$=$4\times 10^7$.}
  \label{fig:survivalprobability}
  \end{subfigure}
\hfill
  \begin{subfigure}[b]{0.4\textwidth}
    \includegraphics[scale=0.8]{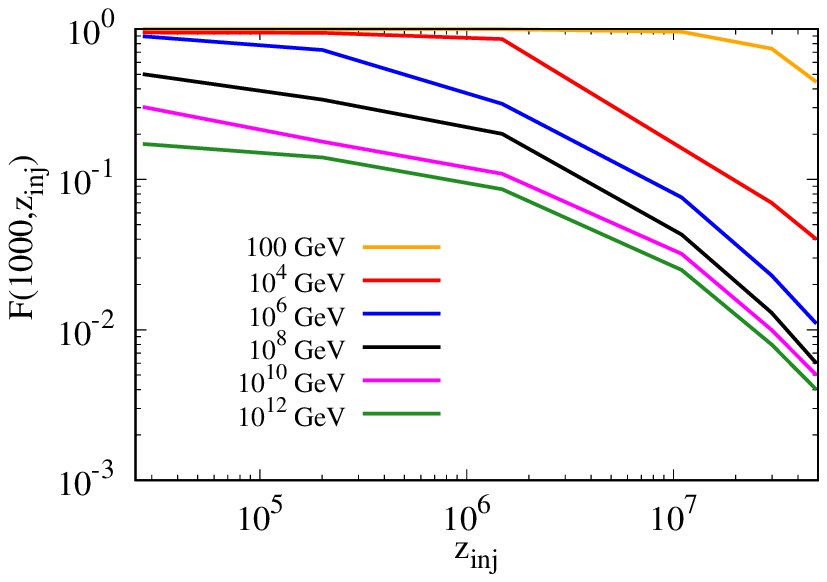}
    \caption{$F(1000,z_{inj})$ for one-time energy injection at $z_{inj}$ for different values of initial neutrino energy. }
    \label{fig:surprob}
  \end{subfigure}  
  \caption{$F(1000,z_{inj})$ for neutrinos as a function of neutrino energy and injection redshift.}
\end{figure}
In Fig. \ref{fig:survivalprobability}, we plot the fraction of injected energy surviving as neutrinos ($F(z,z_{inj})$) at $z<z_{inj}$ for one-time energy injection at $z_{inj}$. The expression for $F(z,z_{inj})$ is given as,
\begin{equation}
F(z,z_{inj})=\frac{\Delta\rho_{\nu}(z)}{\Delta\rho_{\nu}(z_{inj})},
\end{equation}
where $\Delta\rho_{\nu}(z)$ is the surviving neutrino energy density at $z$ and $\Delta\rho_{\nu}(z_{inj})$ is the injected neutrino energy density at $z_{inj}$.
 With increasing energy of injected neutrinos, $F(z,z_{inj})$ decreases as it is more likely for neutrinos to lose their energy after multiple pair-production cycles of quark-antiquarks. This shows up a vertical fall in $F(z,z_{inj})$ as rate of quark-antiquark pair production is much faster than the Hubble rate. In Fig. \ref{fig:surprob}, we plot the extra energy in neutrinos surviving at $z=1000$, $F(1000,z_{inj})$, as a function of $z_{inj}$. The surviving extra energy in neutrinos at $z=1000$ decreases with increase in neutrino energy as well as injection redshifts. For any $z_{inj}$, the efficiency of energy deposition by neutrinos decreases with decreasing redshift as the neutrino energy redshifts and the number density of target particles for collisions also decreases. Thus $F(z,z_{inj})$ approaches an asymptotic constant value for $z<<z_{inj}$. In particular for $z_{inj}>>1000$, we expect $F(z,z_{inj})$ as well as $\Delta N_{eff}$ to be frozen by $z\approx 1000$.    \par
\hspace{1cm} 
\section{CMB spectral distortion constraints}
\begin{figure}
    \includegraphics[scale=1.5]{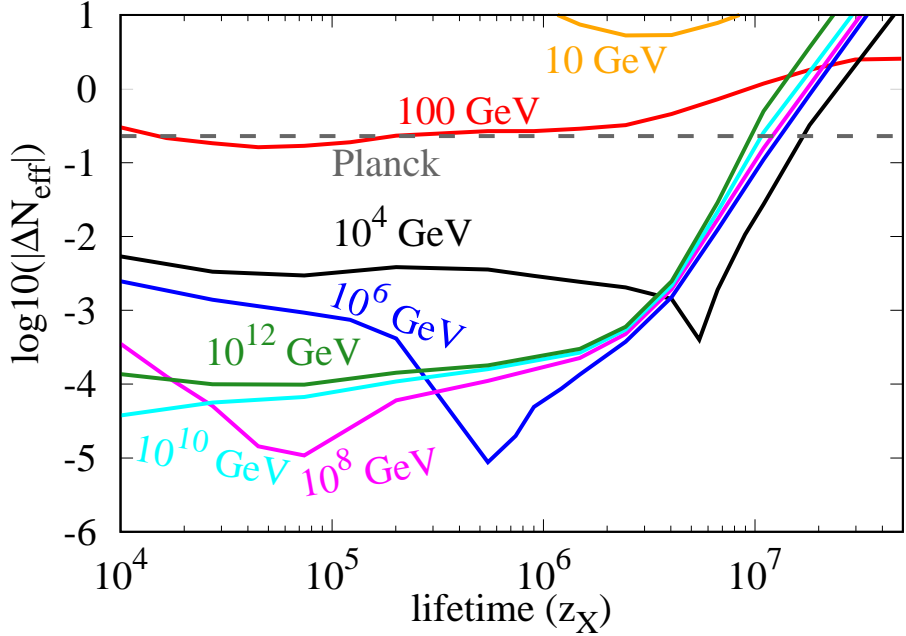}
    \caption{2-$\sigma$ Constraint of $|\Delta N_{\rm{eff}}|$ from CMB spectral distortions ($\mu$-type and $ntr$-type) using COBE-FIRAS data \cite{F1996} as a function of dark matter lifetime for different values of monochromatic neutrino energy. \emph{Planck} constraint \cite{Pl2018} is shown as dashed line. }
    \label{fig:neffconst}
  \end{figure}
  \begin{figure}
  \begin{center}
    \includegraphics[scale=1.0]{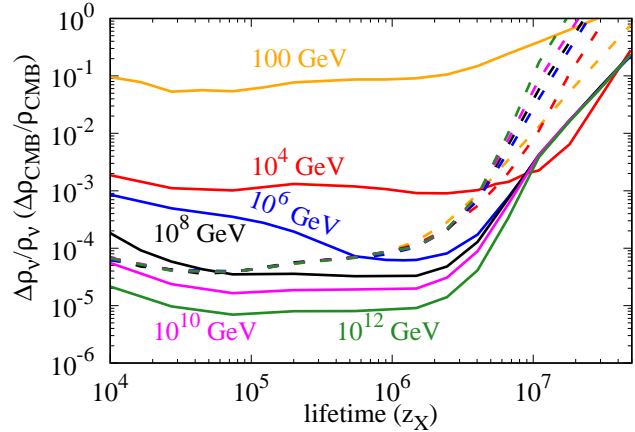}
    \caption{The fractional changes in neutrino energy density $\frac{\Delta \rho_{\nu}}{\rho_{\nu}}$ (solid lines) and CMB energy density $\frac{\Delta \rho_{\rm{CMB}}}{\rho_{\rm{CMB}}}$ (dashed lines) contributing to $|\Delta N_{\rm{eff}}|$ constraints plotted in Fig. \ref{fig:neffconst}.}
    \label{fig:delrho}
    \end{center}
  \end{figure}
  \begin{table}[h!]
\begin{center}
    \begin{tabular}{l|r} 
      lifetime ($z_X$)  & correction factor (C)  \\
      \hline
      $5\times 10^3$ & 1.99 \\
      $6\times 10^3$ & 1.66 \\
      $8\times 10^3$ & 1.21 \\
      $1\times 10^4$ & 1.06 \\
      $2\times 10^4$ & 0.69 \\
      $3\times 10^4$ & 0.63 \\
      $4\times 10^4$ & 0.60 \\     
      $5\times 10^4$ & 0.59 \\
      $6\times 10^4$ & 0.59 \\
      $7\times 10^4$ & 0.60 \\
      $8\times 10^4$ & 0.63 \\
      $9\times 10^4$ & 0.64 \\
      $1\times 10^5$ & 0.67 \\
      $2\times 10^5$ & 0.85 \\
      $3\times 10^5$ & 0.97 \\
      $4\times 10^5$ & 1.009 \\
      $5\times 10^5$ & 1.005 \\
    \end{tabular}
  \end{center}
  \caption{Correction factor for COBE-FIRAS data \cite{F1996} to convert the constraint assuming all energy goes into $\mu$-type distortions to the constraint for $\gtrsim$ 1 GeV $ntr$-type distortions.  }
   \label{tab:table3}
\end{table}
We now use COBE-FIRAS \cite{Cobe1994,F1994,F1996} data to constrain the high energy neutrino
injection. For $z_X>3\times 10^5$, $\mu$-type distortion is a good
approximation\footnote{Note that for decay redshift $z_X$ significant
energy injection will continue for sometime even at $z<z_X$.} and we use the 95 percent $\mu$-distortion limit of
COBE-FIRAS, $\mu < 9\times 10^{-5}$. For $z_X < 3\times 10^{-5}$, $\mu$-type
distortion (or even $i$ or $y$ type distortions) is no longer a good
approximation and we should use the actual spectrum of spectral distortions
  ($ntr$-type distortions) \cite{AK20191}. The constraints for $ntr$-type distortions were derived
 by fitting the $ntr$-type CMB spectral distortions to the COBE-FIRAS data in \cite{AK20191} where it was also shown that when the energy of
  injected particles is $\gtrsim 1 {\rm GeV}$, the CMB spectral distortion
  becomes independent of the energy of the injected particles as well as
  whether the particle is an electron or photon. We will use
  these  $\gtrsim 1 {\rm
    GeV}$ $ntr$-type spectral distortion constraints. This is motivated from the fact that photons/electrons produced
from neutrino/anti-neutrino pair-production have $\gtrsim$ GeV energy
(Fig. \ref{fig:spectrum10^2GeV} and \ref{fig:spectrum10^12GeV}). 

We
  define the  correction factor ($C(z_X)$), a function of $z_X$, as the
  ratio of actual constraint on the amount of energy injected using the
  $\gtrsim 1 {\rm GeV}$ $ntr$-distortion  to the
  constraint we would obtain if we assumed that the energy for $z_X<3\times
  10^{5}$ also created $\mu$-type distortion. Thus a $C>1$ means that the
actual constraints are weaker compared to what we would have inferred if we
assumed that all energy went into $\mu$-type distortions. We give this correction factor 
  in Table \ref{tab:table3} calculated using $ntr$-type distortions constraints of \cite{AK20191}. Note that this correction factor depends on the
  experiment, in particular the sensitivity of different frequency
  channels of the experiment, and is calculated
  for COBE-FIRAS data in Table \ref{tab:table3}. We thus first calculate the
  constraints by assuming that all electromagnetic energy goes into
  $\mu$-type distortion giving $\mu=1.4\epsilon_{\rm SD}$ and then apply
  the correction factor to get the actual constraint.   The correction factor at low redshifts
is  greater than 1 due to reduction of spectral intensity of distortion in
the CMB bands 
as relativistic electrons  boost the CMB photons to high energy tail of the
CMB spectrum and significant amount of energy ends up out of the CMB band. Therefore, higher energy injection
is allowed compared to $\mu$-type distortion. At higher redshifts, the
distorted CMB spectrum partially thermalizes to $i$-type distortion
\cite{Ks2012b,Chluba:2013vsa}, for which energy injection constraint can be
stronger compared to $\mu$-distortion \cite{AK20191}. The correction factor
approaches unity at $z_X\gtrsim 3\times 10^5$.

\par In Fig. \ref{fig:neffconst}, we show 2-$\sigma$ constraints from COBE-FIRAS
on $|\Delta
N_{\rm{eff}}(1000)|$ (from now on, we assume $\Delta N_{\rm{eff}}(z)$ to be
computed at $z=1000$ and we drop the functional dependence on $z$ from now
on) as a function of $z_X$ for different values of initial neutrino
energy, taking into account the correction for the $ntr$-type distortions.
 We also show \emph{Planck} constraint \cite{Pl2018} as dashed line, $|\Delta N_{\rm{eff}}|$  $\sim$ 0.3 at 2-$\sigma$. For 10 GeV neutrinos, the  constraints obtained from CMB spectral distortions are very weak. For neutrinos with energy $\gtrsim 100$ GeV, the spectral distortion constraints become stronger than \emph{Planck}. For 100 GeV neutrinos, the fraction of energy lost to electromagnetic particles through electroweak showers can be 0.1 percent. The fraction of energy lost to electromagnetic particles increases with energy and saturates to about 20 percent at ultra high energies. Since the surviving neutrino energy density dominates the increase in CMB energy density for 100 GeV case, $\Delta N_{\rm{eff}}$ is positive in this case. The kink feature in $10^4-10^8$ GeV neutrinos is due to the change in the sign of $\Delta N_{\rm{eff}}$ as discussed before. The individual contributions, $\frac{\Delta \rho_{\nu}}{\rho_{\nu}}$ and $\frac{\Delta \rho_{\rm{CMB}}}{\rho_{\rm{CMB}}}$ to $\Delta N_{\rm{eff}}$ are plotted in Fig. \ref{fig:delrho} and the kink feature in $\Delta N_{\rm{eff}}$ constraints can be seen in Fig. \ref{fig:delrho} as the crossing of $\frac{\Delta \rho_{\nu}}{\rho_{\nu}}$ and $\frac{\Delta \rho_{\rm{CMB}}}{\rho_{\rm{CMB}}}$ curves. The location of kink or crossover between $\frac{\Delta \rho_{\nu}}{\rho_{\nu}}$ and $\frac{\Delta \rho_{\rm{CMB}}}{\rho_{\rm{CMB}}}$ is a function of neutrino energy. The curves $\frac{\Delta \rho_{\rm{CMB}}}{\rho_{\rm{CMB}}}$ for all values of neutrino energy fall on top of each other at $z_X\lesssim 2\times 10^6$ in Fig. \ref{fig:delrho} as most of electromagnetic energy deposited to the CMB shows up as spectral distortion and is constrained to be of the order $10^{-4}$ by COBE-FIRAS \cite{Cobe1994,F1996}. For $z_X\gtrsim 2\times 10^6$, the   electromagnetic energy deposited at $z\gtrsim 2\times 10^6$ does not produce a spectral distortion and is, therefore, not constrained by the data. The $\frac{\Delta \rho_{\rm{CMB}}}{\rho_{\rm{CMB}}}$ curves in Fig. \ref{fig:delrho} diverge from each other as the amount of electromagnetic energy deposited in the CMB at $z\lesssim 2\times 10^6$ for neutrino energy injection at $z\gtrsim 2\times 10^6$, and therefore the constraint from CMB spectral distortions, is a function of neutrino energy. Neutrinos injected at high redshifts will rapidly pair produce, giving rise to a broad low energy neutrino spectrum below the pair production threshold erasing any information of original neutrino energy. This makes the $\frac{\Delta \rho_{\nu}}{\rho_{\nu}}$ curves converge to each other for high initial neutrino energies.  With increase in neutrino energy, the threshold of pair production of quark-antiquark pairs (also true for electron-positron pairs but is unimportant as pair production rate of $e^-e^+$ is much slower compared to the Hubble rate (Fig. \ref{fig:rate})) occurs at lower redshifts. The higher energy ($\gtrsim 10^6$ GeV) cases are all pair production dominated at $z \lesssim 2\times 10^6$. Therefore, $\Delta N_{\rm{eff}}$ is negative for all redshifts for these cases. For pair production to $e^-e^+$, all of the neutrino energy and for quarks almost 50 percent of the neutrino energy is released as electromagnetic particles. The secondary neutrinos can themselves pair produce again and again, the probability of which increases with increase in incident neutrino energy. This increases the probability of electromagnetic energy release which reduces the energy density in surviving neutrinos or $\frac{\Delta \rho_{\nu}}{\rho_{\nu}}$ .  \par
\hspace{1cm}   
\begin{figure}
\begin{center}
    \includegraphics[scale=1.0]{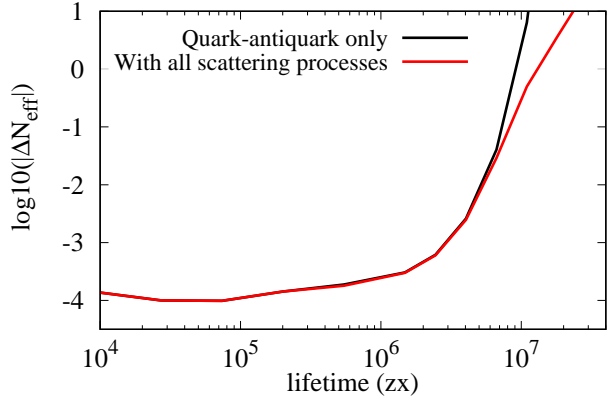}
    \caption{2-$\sigma$ Constraints on $|\Delta N_{\rm{eff}}|$ from CMB spectral distortions including only quark-antiquark pair production and including quark-antiquark, electron-positron pair production and neutrino-nucleon scattering. Energy of injected neutrino is $10^{12}$ GeV.}
    \label{fig:neffconst3}
    \end{center}
  \end{figure}
In Fig. \ref{fig:neffconst3}, we compare the constraints on $\Delta N_{\rm{eff}}$ including only quark pair production process with the case when we include all scattering processes i.e. including quark-antiquark, electron-positron pair production, and neutrino-nucleon inelastic scattering. From Fig. \ref{fig:rate}, we can see that the  quark pair production rate is much faster than the Hubble rate. Therefore, electromagnetic release from neutrinos is instantaneous just like pure electromagnetic energy injection. Hence, the constraints with only quark-antiquark pair production track the $\mu$-visibility curve. The low energy surviving neutrinos below the quark pair production threshold can deposit their energy at much slower rate compared to the Hubble rate. Therefore, with these additional processes, we can probe higher redshifts as shown in Fig. \ref{fig:neffconst3}. At $z\lesssim 2\times 10^6$, we can ignore the neutrino-nucleon scattering and $e^-e^+$ pair production, since quark pair production drives the $\Delta N_{\rm{eff}}$ constraints as long as the energy of injected neutrinos exceeds the pair production threshold.  \par
\hspace{1cm}
\section{Constraints from abundance of light elements}
\begin{table}[h!]
\begin{center}
    \begin{tabular}{l|c|r} 
      Elements & theoretical value(1$\sigma$)  & observational value(1$\sigma$)  \\
      \hline
      $n_{^2{\rm H}}/n_{\rm H}$ & $(2.58\pm 0.13)\times 10^{-5}$ \cite{CFOY2016} & $(2.53\pm 0.04)\times 10^{-5}$  \cite{CFOY2016}\\
     $Y_p$ & $0.24709\pm 0.00025$ \cite{CFOY2016} & $0.2449\pm 0.0040$ \cite{AOS2015}\\
      $n_{^3{\rm He}}/n_{\rm H}$ & $(10.039\pm 0.090)\times 10^{-6}$ \cite{CFOY2016} & $1.5\times 10^{-5}$ (2$\sigma$ upper limit) \cite{BRB2002}\\
    \end{tabular}
  \end{center}
  \caption{Theoretical and observational  bounds on light element abundance that are used in this work for deuterium ($n_{^2{\rm
        H}}$) and helium-3 number density ($n_{^3{\rm He}}$)  with respect to hydrogen number density ($n_{\rm H}$). Helium-3 and Deuterium over-production provide stronger constraints compared to change in Helium-4 ($^4{\rm He}$) mass fraction ($Y_P$), where $Y_p$ is the ratio of helium mass density to the total mass density of hydrogen and helium.}
   \label{tab:table2}
\end{table}
High energy photons, in the electromagnetic shower produced by injected neutrinos above helium and deuterium photo-dissociation thresholds (19.81 and 2.2 MeV respectively), can dissociate helium-4 producing helium-3 and deuterium or destroy deuterium, thus changing the primordial abundance of these light nuclei. We can use the theoretical and observational bounds on the abundance of helium-3 and deuterium to constrain the energy injection scenarios. The bounds on abundance used in this calculation are given in Table \ref{tab:table2} and the procedure to obtain the constraints is detailed in \cite{AK20192}. 
  To obtain BBN constraints from electromagnetic energy injection, we have to evolve the  electromagnetic cascade along with the abundance of primordial elements, which are being created and destroyed, in the expanding universe \cite{KM1995,PS2015,KKMT2018,HSW2018,FMW2019,AK20192}. Photons are evolved with Compton scattering, $e^-e^+$ pair production on the CMB photons and background electrons and nuclei, photon-photon elastic scattering, and photo-dissociation of elements. There is a competition between photodissociation and other processes which degrade the injected photon energy. When the energy of degraded photons fall below $\sim$ 20 MeV, there is no photo-dissociation of helium-4 anymore. The constraints obtained from BBN for high energy photons  and electrons ($>$GeV) are essentially universal and do not depend upon the energy of injected photon and electron \cite{KM1995}. This is because high energy photon scattering processes such as $e^-e^+$ pair production on the CMB and photon-photon elastic scattering are extremely efficient in processing the high energy photons and creating a broad low energy spectrum, irrespective of the energy of the  original photon. Recently, it was shown that this may not be true for 10-100 MeV electromagnetic particles and the constraints are non-universal \cite{PS2015,HSW2018,FMW2019,AK20192}. Here, we do a simplified analysis in the on-the-spot approximation which means that all the electromagnetic energy injected at a redshift either creates or destroys elements or degrades to sub-MeV energy at that redshift (see Appendix \ref{app:bbn} for a discussion on comparison of on-the-spot approximation and the full calculation).  We also ignore the shape of the electromagnetic spectrum above 2.2 MeV and assume all particles to have energy $\gtrsim$GeV. This is a good approximation as most of the secondary photon/electron have energy greater than GeV as can be seen in Fig. \ref{fig:spectrum10^2GeV} and \ref{fig:spectrum10^12GeV}. 
\par
\hspace{1cm}  
    We first tabulate what fraction of a $\sim$GeV photon's or electron's energy is used up as creation and destruction of primordial elements as a function of redshift. We then use this tabulated data with the electromagnetic energy released from the neutrino cascade to get the change in the primordial nuclear abundance as a function of redshift. We find that the strongest constraints come from creation of helium-3 by destruction of helium-4 as the cross-section for photo-dissociation of helium-4 to helium-3 is an order of magnitude higher compared to dissociation of helium-4 to deuterium. There are however systematic uncertainties in astrophysical observation of helium-3 \cite{HKKM1999,KKMT2018}. Constraints from deuterium destruction are weak due to its low abundance ($\sim 10^{-5}$ times the number density of helium-4) and we do not consider them in this paper.\par
  \hspace{1cm}
  \begin{figure}
    \includegraphics[scale=1.5]{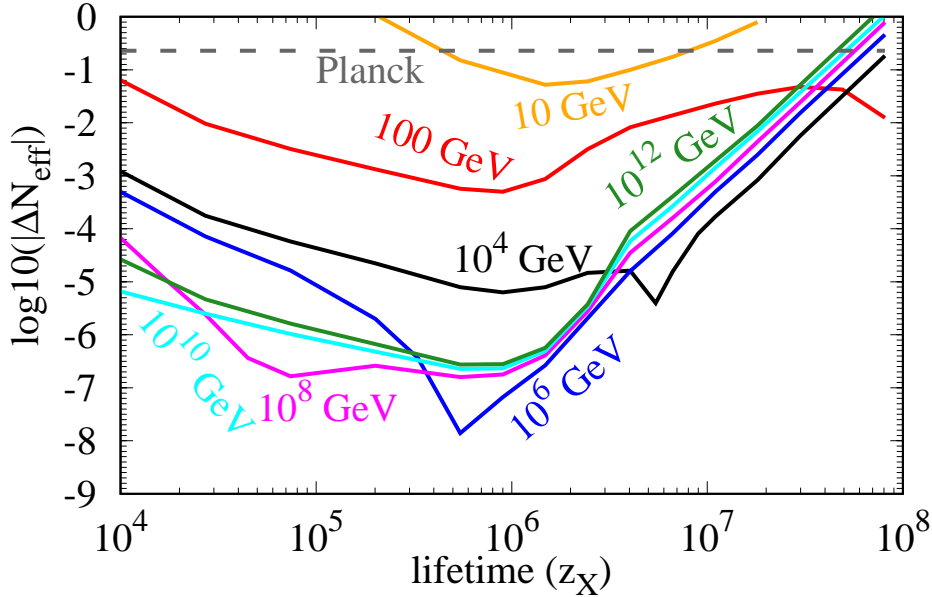}
    \caption{2-$\sigma$ Constraint of $|\Delta N_{\rm{eff}}|$ from change in primordial nuclear abundance from destruction of helium-4 into helium-3 as a function of dark matter lifetime with different monochromatic injected neutrino initial energy.}
    \label{fig:neffconst4}
  \end{figure}
  \begin{figure}
    \includegraphics[scale=1.5]{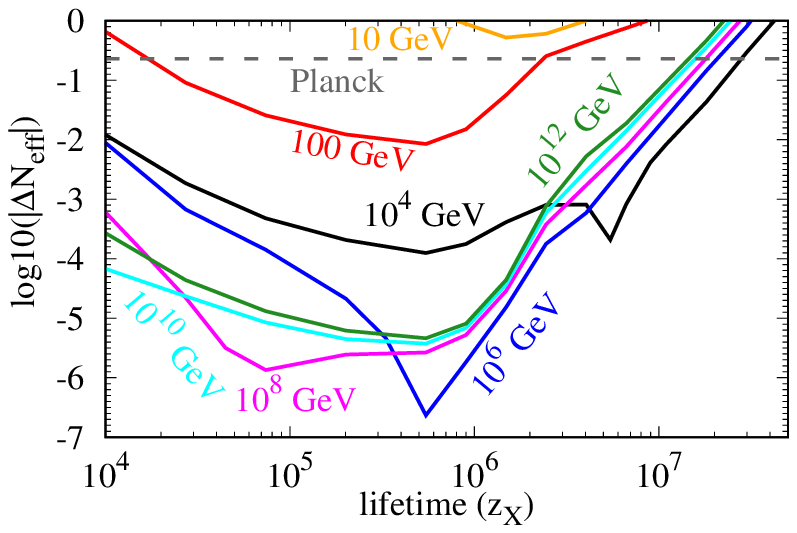}
    \caption{2-$\sigma$ Constraint on $|\Delta N_{\rm{eff}}|$ from change in primordial nuclear abundance of helium-3 by destruction of helium-4 into deuterium as a function of dark matter lifetime with different values of monochromatic injected neutrino initial energy.}
    \label{fig:neffconst41}
  \end{figure}
 \begin{figure}
\begin{center}
    \includegraphics[scale=1.0]{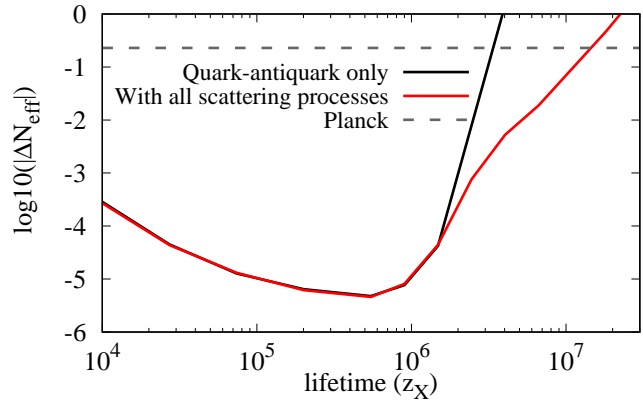}
    \caption{2-$\sigma$ Constraints on $|\Delta N_{\rm{eff}}|$ from abundance of deuterium (overproduction of deuterium from destruction of helium) including only quark-antiquark pair production and including quark-antiquark as well as electron-positron pair production and neutrino-nucleon scattering. Energy of injected neutrino is $10^{12}$ GeV. }
    \label{fig:neffconst5}
    \end{center}
  \end{figure}
  \begin{figure}[!tbp]
  \begin{subfigure}[b]{0.4\textwidth}
    \includegraphics[scale=0.8]{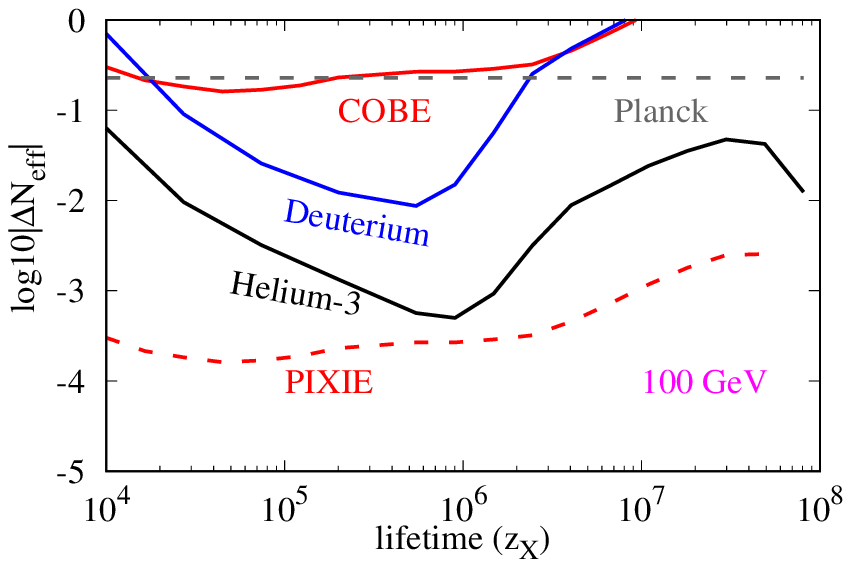}
    \label{fig:100gev}
  \end{subfigure}
  \hfill
  \begin{subfigure}[b]{0.4\textwidth}
    \includegraphics[scale=0.8]{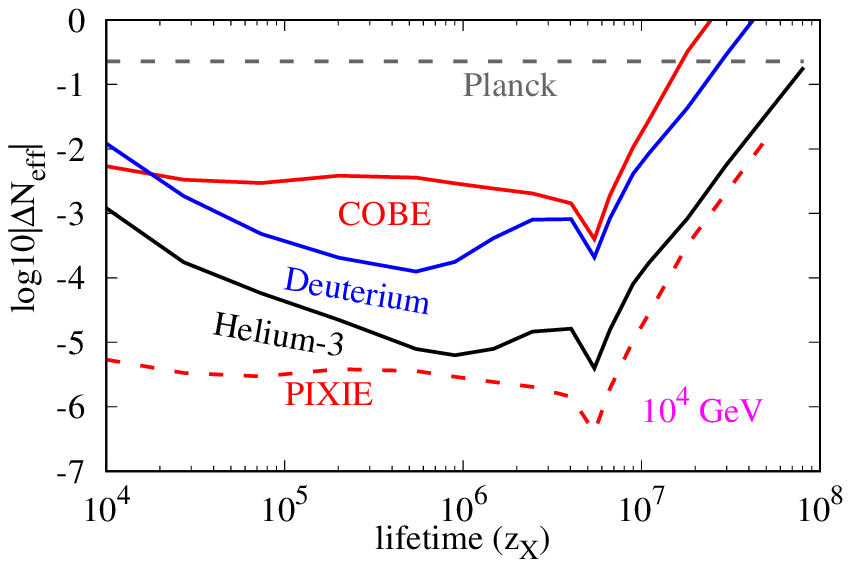}
    \label{fig:10^4gev}
  \end{subfigure}
  \begin{subfigure}[b]{0.4\textwidth}
    \includegraphics[scale=0.8]{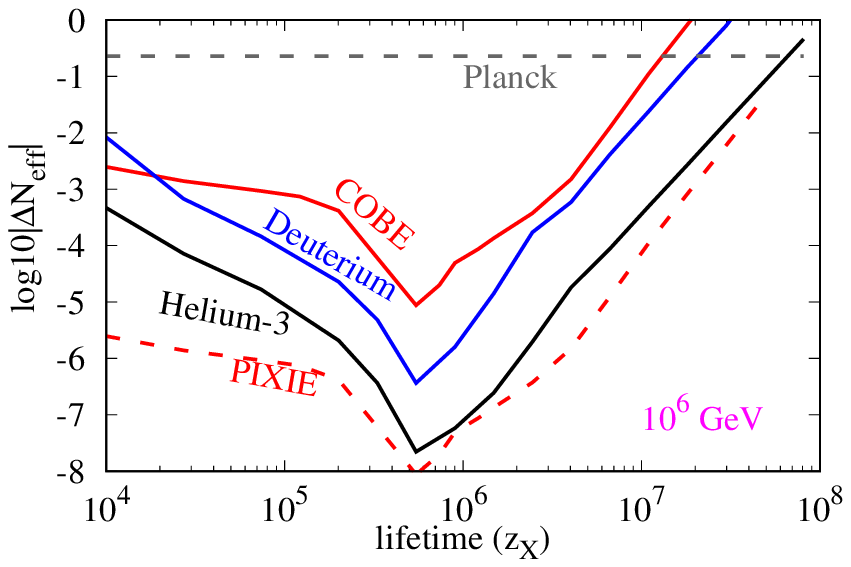}
    \label{fig:10^6gev}
  \end{subfigure}
  \hfill
  \begin{subfigure}[b]{0.4\textwidth}
    \includegraphics[scale=0.8]{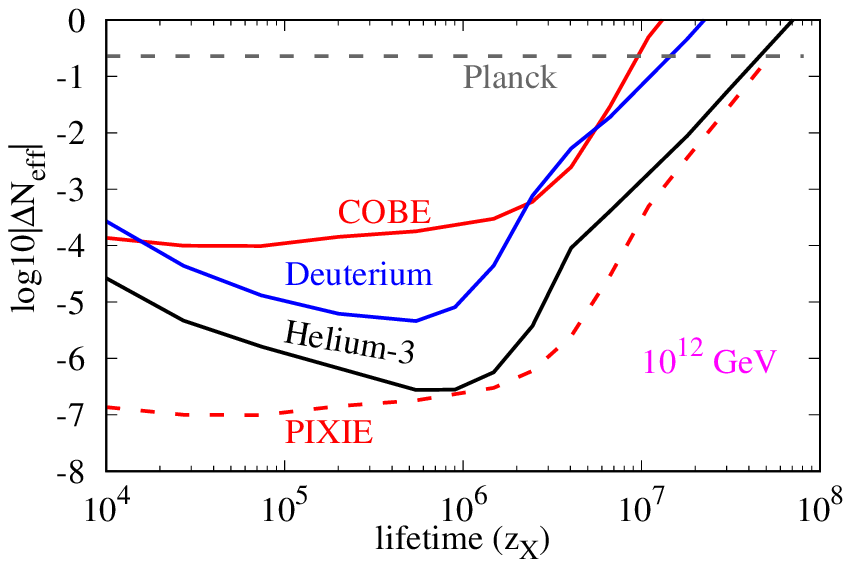}
    \label{fig:10^12gev}
  \end{subfigure}
    \caption{2-$\sigma$ Constraints on $|\Delta N_{\rm{eff}}|$ from COBE and abundance of BBN elements for (a) 100 GeV (b) $10^4$ GeV (c) $10^6$ GeV (d) $10^{12}$ GeV neutrino energy. BBN constraints are from helium-4  destruction to helium-3 and deuterium. Also shown are projection for future experiment PIXIE \cite{Pixie2011}.}
    \label{fig:neffconst6}
  \end{figure}
  \begin{figure}[!tbp]
  \begin{subfigure}[b]{0.4\textwidth}
    \includegraphics[scale=0.8]{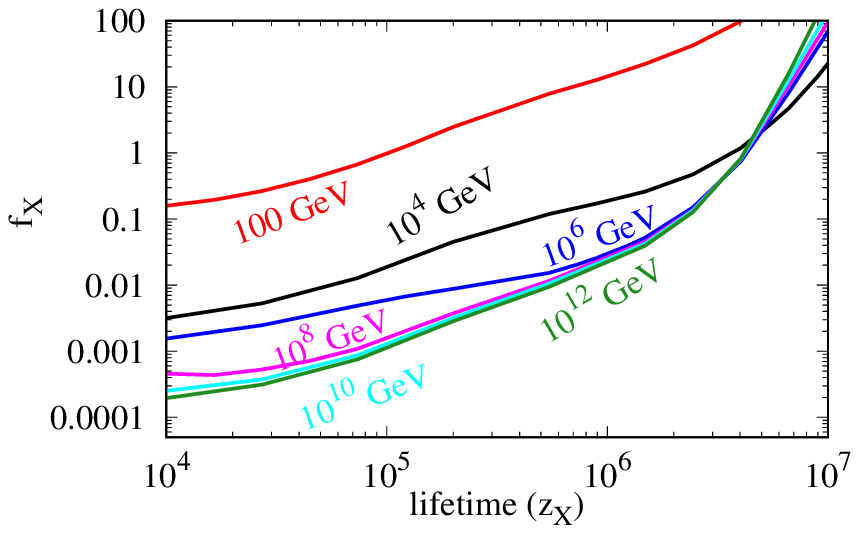}
    \caption{CMB spectral distortion}
    \label{fig:fxspecdist}
  \end{subfigure}
  \hfill
  \begin{subfigure}[b]{0.4\textwidth}
    \includegraphics[scale=0.8]{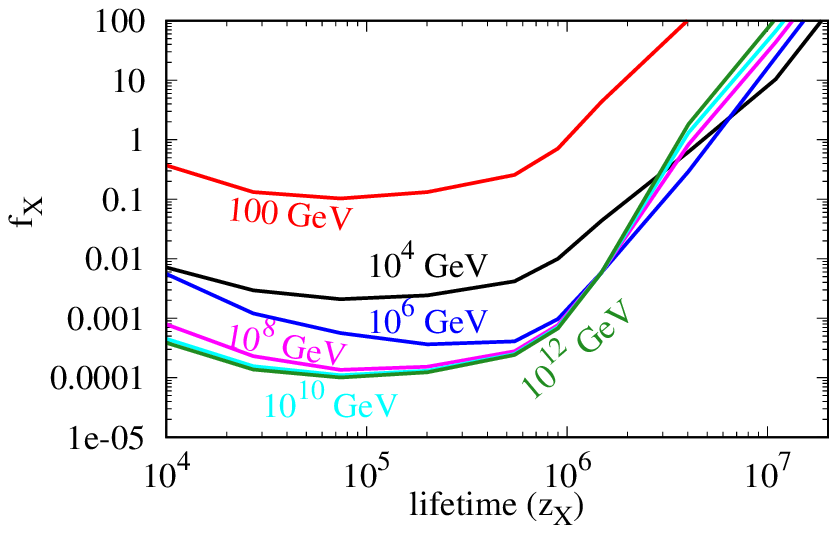}
    \caption{Deuterium abundance}
    \label{fig:fxbbndeu}
  \end{subfigure}
  \caption{2-$\sigma$ constraints on fraction of decaying dark matter $f_X$, as a function of lifetime from (a) CMB spectral distortion (b) deuterium abundance (over-production of deuterium from destruction of helium-4). The labels correspond to the energy of initial neutrino injected. }
  \label{fig:fxconst}
  \end{figure}
  In Fig. \ref{fig:neffconst4} and \ref{fig:neffconst41}, we plot constraints on $\Delta N_{\rm{eff}}$ allowed by abundance of primordial elements. We see similar patterns as in the case of spectral distortions. The constraints initially get stronger with increasing redshifts. Electrons and positrons in the electromagnetic cascade can boost the CMB photons to higher energy at higher redshifts as the average energy of CMB photons are higher. Thus, more photons are available above the photo-dissociation threshold of helium-4. At still higher redshifts, constraints weaken as the injected photons can efficiently pair produce $e^-e^+$ on the CMB photons. Once the electron-positron pair production threshold of energetic photons on the CMB photons is of similar energy as compared to the photo-dissociation threshold of helium-4, the photons are degraded efficiently to sub-MeV energy instead of destroying helium, resulting in weakening of constraints. This can be clearly seen in Fig. \ref{fig:neffconst5}. With quark pair production only, the constraints weaken much faster as compared to the case when we consider the lower energy scattering processes of neutrinos. The $\Delta N_{\rm{eff}}$ constraints obtained for COBE \cite{F1996}, abundance of light elements, and projection from PIXIE \cite{Pixie2011} are summarized in Fig. \ref{fig:neffconst6}. In Fig. \ref{fig:fxconst}, we plot the 2-$\sigma$ constraint for decaying dark matter abundance for the neutrino decay channel as a function of its lifetime. Since energy injection from dark matter decay scales with redshift as $(1+z)^3$ while energy density of CMB goes as $(1+z)^4$, constraint on $f_X$ from CMB spectral distortion scales as $(1+z)$ for $z\lesssim 10^6$. The abundance of light elements scales as $(1+z)^3$, therefore, constraints on $f_X$ in this case are flatter with respect to redshift of injection for $z\lesssim 10^6$. 
  \section{Conclusion}
  In this work, we have obtained constraints on injection of high energy neutrinos in the early universe using dark matter decay as an example and the resulting free streaming degrees of freedom at recombination epoch parameterized by $\Delta N_{\rm{eff}}$. High energy neutrinos (anti-neutrinos) can deposit a fraction of their  energy as electromagnetic energy by pair producing standard model particles on background anti-neutrinos (neutrinos), electroweak showers from electroweak bremsstrahlung during decay or inelastic neutrino-nucleon scattering. The secondary electromagnetic energy injection can be constrained by CMB spectral distortions or light element abundance from BBN which indirectly puts constraints on the allowed fractional energy density of injected non-thermal neutrinos and the surviving extra neutrino energy  density during recombination ($\Delta N_{\rm{eff}}$). We have shown that the CMB spectral distortion and the abundance of light elements strongly limit which sources or new physics can contribute to $N_{\rm{eff}}$. In particular, we get constraints which are several orders of magnitude stronger compared to current $Planck$ constraints on $N_{\rm{eff}}$ for models where the injected neutrinos have energy $\gtrsim$ 100 GeV. We, therefore, rule out new physics which injects $\gtrsim $100 GeV neutrinos in the early universe after neutrino decoupling as a significant source of deviation of $N_{\rm{eff}}$ from the standard $\Lambda$CDM value.  To our knowledge, this is the first calculation which evolves $N_{\rm{eff}}$ from particle cascades in the early universe, does not assume $N_{\rm{eff}}$ to be a constant, and takes into account  interaction of neutrinos with the background particles. We have also shown that with neutrino, we can probe deeper in redshift as compared to pure electromagnetic energy injection as the  surviving neutrinos from $z\gtrsim 2\times 10^6$ can deposit a fraction of their energy as electromagnetic energy at $z\lesssim 2\times 10^6$. This opens up a new window to study energy injection history beyond CMB black body surface allowing us to peek into the CMB black body photosphere or the thermalization epoch.      
 \section{Acknowledgements}
 We acknowledge the use of computational facilities of Department of
 Theoretical Physics at Tata Institute of Fundamental Research,
 Mumbai. This work was supported by Max Planck Partner Group for cosmology of Max Planck
 Institute for Astrophysics Garching at 
 Tata Institute of Fundamental Research funded by
 Max-Planck-Gesellschaft. We acknowledge support of the Department of Atomic Energy, Government of India, under project no. 12-R\&D-TFR-5.02-0200. SKA is grateful for financial support from the Royal Society and Prof. Jens Chluba for the invitation to University of Manchester, during which a part of this work was done.  
\appendix
\section{Computation of particles cascade}\label{app:algo}
The evolution of particle spectra can be written as \cite{KK2008,Slatyer:2009yq},
\begin{equation}
\frac{\partial N^{\alpha}_a}{\partial t}=-\left(\sum\limits_{\beta < \alpha}\sum\limits_b P^{\alpha \beta}_{ab}N^{\alpha}_a\right)+\left(\sum\limits_{\beta > \alpha}\sum\limits_b P^{\beta \alpha}_{ba}N^{\beta}_b\right)+S^{\alpha}_a,
\label{evolution}
\end{equation}
   where $N^{\alpha}_a, N^{\beta}_b$ are the number of particles of type $a$, $b$ (where $a$ and $b$ can be neutrinos (anti-neutrinos), photons or electrons (positrons)) in the bin denoted by $\alpha$, $\beta$ with energy $E_{\alpha}$ and $E_{\beta}$ respectively, $P^{\alpha \beta}_{ab}$ is the rate of particle of type $a$ to transfer from bin $\alpha$ to $\beta$ and particle type $b$ with $E_{\beta}<E_{\alpha}$ in a timestep, $S^{\alpha}_a$ is a source or sink function which can be non-zero for particle of type $a$ injection or destruction in bin $\alpha$ at a particular timestep. For neutrinos above quark pair production threshold, $P^{\alpha \beta}_{ab}$ for $a$=neutrino and $b$=(neutrino, photon, electron) can be obtained directly from \textbf{PYTHIA} data of \cite{CCHHKPRSS2011}. The authors in \cite{CCHHKPRSS2011} provide spectrum of secondary neutrinos, photons and electrons after hadronization of quark-antiquark pair from neutrino-antineutrino pair-production. For neutrino-nucleon scattering and electromagnetic processes, $P^{\alpha \beta}_{ab}$ can be computed from the cross-section of these processes. The expression for $P^{\alpha \beta}_{ab}$ is given by,
\begin{equation}
   P^{\alpha \beta}_{ab}=\sum n_T\rm{c}\frac{d\sigma^{\alpha\beta}_{ab}}{dE^{\beta}}\Delta E^{\beta},
   \end{equation}
   where $n_T$ is the number density of target particles for a particular scattering process, $\frac{d\sigma^{\alpha\beta}_{ab}}{dE^{\beta}}$ is the differential cross-section for a particle of type $a$ in bin $\alpha$ to scatter to bin $\beta$ as a particle of type $b$. The cross-section for neutrino-nucleon scattering and electromagnetic processes, used in this work, can be found in \cite{GQHS1996} and \cite{AK2018} respectively and references therein. For spectral distortion calculations, the energy of electromagnetic particles produced in a time step from quark hadronization and neutrino-nucleon scattering are just summed up and there is nothing else to be done for these particles. For BBN elements calculations, we ignore the time evolution of electromagnetic particles which just means putting left hand side of Eq. \ref{evolution} for these particles to be zero. This is the on-the-spot approximation which assumes that all the injected electromagnetic energy is deposited on time scales smaller than the Hubble time. By solving the resulting algebraic equations \cite{KK2008,Slatyer:2009yq,AK2018}, we obtain the secondary electromagnetic spectrum and energy deposited as photo-dissociation of nuclei, from interaction of $\sim$ GeV energy electromagnetic particles with background photons, electrons and nuclei.  
   
 \section{Comparison of on-the-spot and full cascade calculations on constraints of abundance of light elements}\label{app:bbn}
  \begin{figure}
\begin{center}
    \includegraphics[scale=1.0]{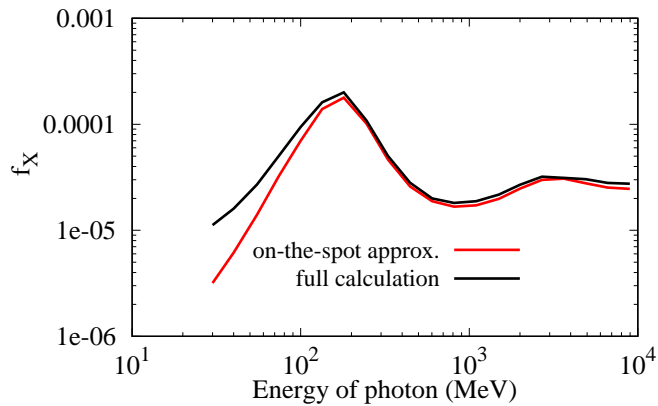}
    \caption{2-$\sigma$ constraint on $f_X$ from helium-3 abundance for dark matter decay to monochromatic photon pairs as a function of photon's energy. The lifetime of dark matter is $z_X=20000.$ }
    \label{fig:bbnenergy}
    \end{center}
  \end{figure}
  We show the comparison of the simplified calculation with on-the-spot approximation with the full cascade calculation in Fig. \ref{fig:bbnenergy}. We plot the constraints obtained from helium-3 abundance for dark matter decay to monochromatic photon pairs with $z_X=20000$. For photons with energy $\gtrsim$GeV, on-the-spot approximation is very good with only slightly stronger constraints compared to the full calculation.
  \newpage
\bibliographystyle{unsrtads}
\bibliography{neutrino} 
\end{document}